\documentclass[10pt,twocolumn,letterpaper,groupedaddress]{revtex4}

\usepackage{multibbl}
\usepackage[pdftex]{graphicx}
\usepackage{amsmath,SIunits}
\usepackage[latin1]{inputenc}
\usepackage{latexsym,stmaryrd}
\usepackage[sort&compress]{natbib}
\usepackage[pdftex]{graphicx}
\usepackage{color}
\bibpunct{[}{]}{,}{n}{}{,}

\makeatletter
\newcommand*{\citenst}[2][]{%
  \begingroup
  \let\NAT@mbox=\mbox
  \let\@cite\NAT@citenum
  \let\NAT@space\NAT@spacechar
  \let\NAT@super@kern\relax
  \renewcommand\NAT@open{[}%
  \renewcommand\NAT@close{]}%
  \citep{#2}%
  \endgroup
}
\makeatother

\begin{document}

\title{Physics of quantum light emitters in disordered photonic nanostructures}

\author{P. D. Garc\'{i}a} \email{david.garcia@icn2.cat} \affiliation{Catalan Institute of Nanoscience and Nanotechnology (ICN2), CSIC and The Barcelona Institute of Science and Technology, Campus UAB, Bellaterra, 08193 Barcelona, Spain} \author{P. Lodahl} \email{lodahl@nbi.ku.dk} \homepage{http://www.quantum-photonics.dk/} \affiliation{Niels Bohr Institute, University of Copenhagen, Blegdamsvej 17, DK-2100 Copenhagen, Denmark}
\date{\today}

\small

\begin{abstract} Nanophotonics focuses on the control of light and the interaction with matter by the aid of intricate nanostructures.\ Typically, a photonic nanostructure is carefully designed for a specific application and any imperfections may reduce its performance, i.e., a thorough investigation of the role of unavoidable fabrication imperfections is essential for any application.\ However, another approach to nanophotonic applications exists where fabrication disorder is used to induce functionalities by enhancing light-matter interaction.\ Disorder leads to multiple scattering of light, which is the realm of statistical optics where light propagation requires a statistical description.\ We review here the recent progress on disordered photonic nanostructures and the potential implications for  quantum photonics devices.
\end{abstract}

\maketitle

\section{I\lowercase{ntroduction: focus and scope of this} R\lowercase{eview}.}

By harnessing the interaction between light and matter, new technologies have arisen ranging from incandescent light emission to single-photon sources.\ Therefore, controlling how light is emitted and absorbed is at the forefront of modern photonics and is at the heart of, e.g., energy harvesting or photonic quantum technology.\ It is an essential insight that the emission of light can be controlled by tailoring the material environment surrounding the emitter~\cite{Purcell}.\ Consequently, different approaches have been taken to engineer photonics at the nanoscale primarily by using ordered photonic structures such as high finesse Fabry-Perot microcavities, whispering gallery resonators, or microcavities and waveguides based on photonic crystals~\cite{Vahala}.\ An overall vision of the research field is to construct complex photonic circuits of basic and highly-controlled elements for both classical~\cite{Joannopoulos} and quantum~\cite{Vucovick} applications.\ However, such nanostructures are sensitive to small fabrication imperfections~\cite{Akahane,Savona_optimization} and calls for a systematic assessment of the role of disorder on the device performance in any application.\ Interestingly, disorder may also be employed in some cases as a resource, notably leading to phenomena such as efficient light trapping~\cite{ALlight,Luca} and random lasing~\cite{Lawandy,Wiersma1}.

\begin{figure}[t!] \includegraphics[width=\columnwidth]{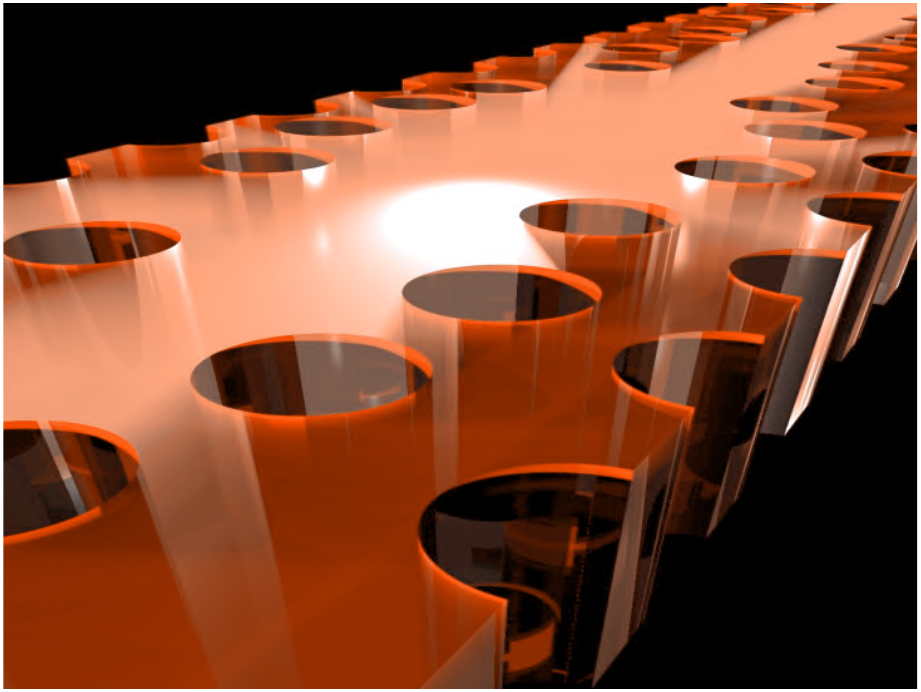} \caption{\label{1} \textbf{Illustration of one-dimensional Anderson localization in a disordered photonic-crystal waveguide.}
Light emitted from internal light sources inside the waveguide may be trapped due to random scattering from disorder leading to the formation of cavities.\ Illustration made by S. Stobbe.} \end{figure}

As a consequence, an increasing interest in unconventional photonic nanostructures exploiting disorder has emerged giving rise to a new field of research referred to as \emph{disordered photonics}~\cite{Wiersma2,Koenderink,PCdisorder}.\ Significant attention has been devoted to explore the role of disorder in photonic-crystal waveguides~\cite{Lalanne1,Vollmer,Luca,Hughes,Savona}, e.g., to unravel their performance limitations.\ Interestingly, the performance of disordered nanostructures has shown to be surprisingly good.\ One interesting example was reported by the group of F. Vollmer, who observed surprisingly narrow cavity resonances in the slow-light regime of photonic-crystal waveguides notably outperforming the quality (Q)-factor achieved in their engineered nanocavities~\cite{Vollmer}.\ These high-Q cavity modes were unequivocally attributed to unavoidable disorder featuring in certain frequency ranges of state-of-the-art photonic-crystal waveguides.\ These modes are due to one-dimensional Anderson localization of light: the coherent scattering of light by the imperfections in the waveguide leads to the spontaneous formation of random cavities.\ Anderson-localized photonic-crystal waveguide cavities are illustrated in Fig.~\ref{1}.

Here, we review recent experiments on the physics of light emitters in disordered media.\ We limit the scope to primarily consider photonic crystals where disorder may deliberately be added as a perturbation to the existing lattice.\ Far-field photoluminescence measurements are used to probe the statistical properties of these structures and cavity quantum-electrodynamic experiments reveal the strong interaction between the internal emitters and Anderson-localized modes.\ In addition, we briefly discuss fully random (amorphous) structures with embedded emitters. Finally, we outline some of the next research challenges and opportunities in the research field with the prospects of scaling up and manipulating Anderson-localized modes in more complex architectures.

\section{T\lowercase{he role of disorder in photonic structures revealed by internal emitters}.}

\begin{figure}[t] \includegraphics[width=\columnwidth]{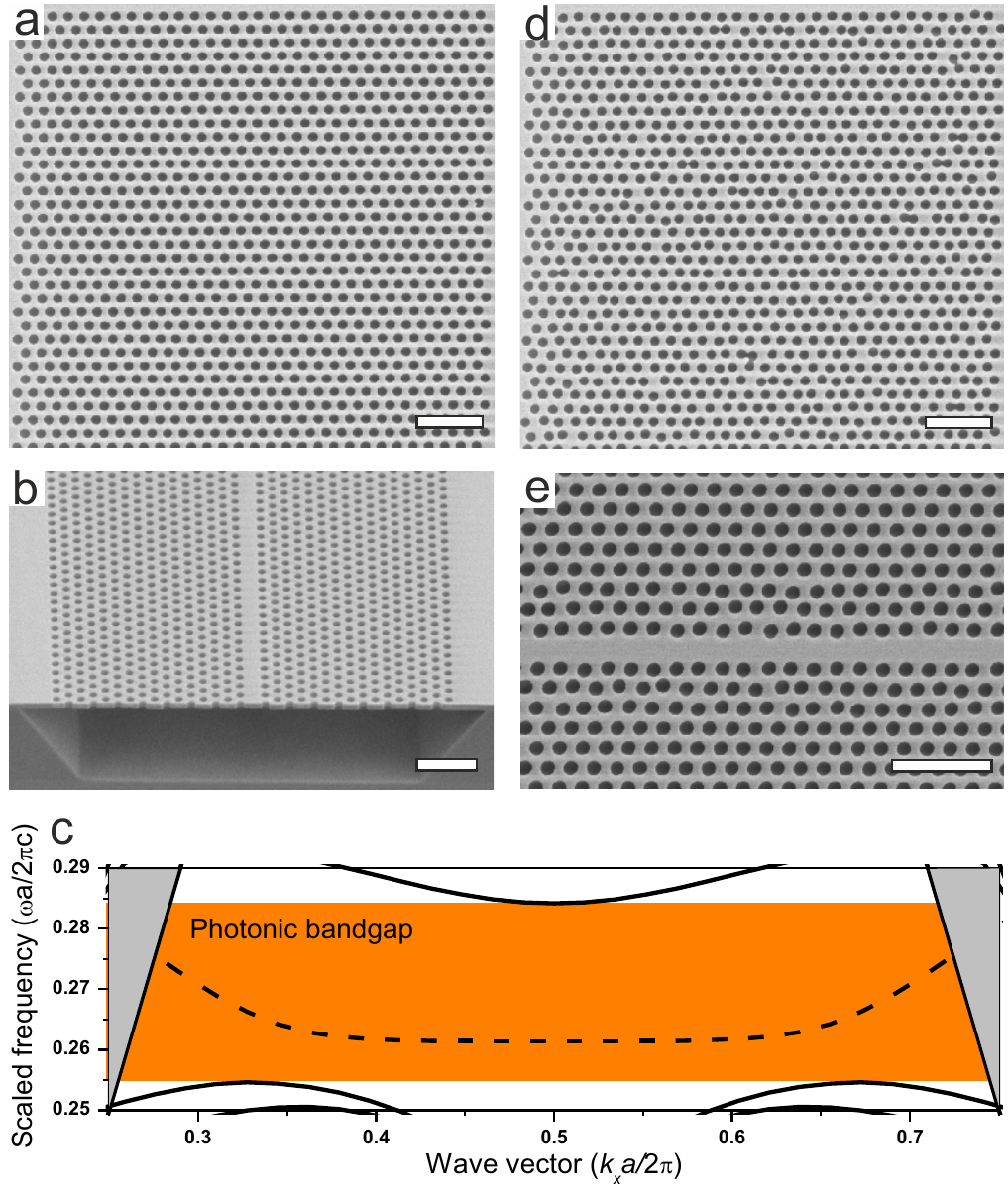} \caption{ \label{2}
Scanning electron micrographs of a photonic-crystal membrane featuring a 2D photonic band gap (\textbf{a}) and a photonic-crystal waveguide obtained by leaving out a row of holes in the photonic crystal (\textbf{b}).\ \textbf{c} Dispersion relation of TE-like modes of a photonic-crystal waveguide where the dashed curve corresponds to an even-parity guided mode in the waveguide.\ The lattice constant is  $a=260\,\text{nm}$ and hole radius $r=0.29a.$\ The orange-shaded region around the mode outlines the photonic bandgap region.\ \textbf{d} and \textbf{e} display scanning electron micrograph images (top view) of  disordered photonic crystals and photonic-crystal waveguides where the positions of the holes are randomized with a standard deviation of $\sigma = 0.06a$.\ The scale bar in all the micrographs corresponds to $1\,\micro\text{m}$.\ Image \textbf{b} from Ref.~\cite{Luca}.} \end{figure}

In this section we introduce and review disordered photonic crystals and photonic-crystal waveguides that can be probed by embedding internal quantum emitters inside the structures.  Figure~\ref{2} \textbf{a} and \textbf{b} show a photonic crystal with a photonic band gap enabling the control of spontaneous emission and a photonic-crystal waveguide that can be used to efficiently route photons to a single propagating mode.\ These particular samples were obtained by etching a triangular lattice of holes in a semiconductor membrane of GaAs leading to three-dimensional confinement of light.\ In photonic crystals, the modulated refractive index shapes the electric field locally and the corresponding photon dispersion relation can be engineered featuring tailored bands and band gaps, cf.~Fig.~\ref{2}\textbf{c} for an example of the dispersion relation of a photonic-crystal waveguide.\ In photonic-crystal waveguides and cavities, light and matter can be very efficiently interfaced by enhancing the coupling to one preferential guided or localized mode while simultaneously inhibiting the coupling to all leaky modes.\ This allows increasing the available local density of optical states (LDOS) and therefore enhancing the emission rate of quantum light emitters~\cite{reviewMod}, which is the main motivation for using these type of structures for quantum photonics.

Photonic-crystal membranes can be fabricated by electron-beam lithography patterning followed by etching on semiconductor wafers containing quantum light sources such as quantum dots or quantum wells that thereby are incorporated directly inside the nanostructures.\ As for any fabrication method, this unavoidably introduces entropy, i.e., statistical errors in the form of variations relative to the ideal target structure due to a finite fabrication precision.\ This manifests itself, e.g., in statistical size and shape variations as well as roughness of the holes of the photonic crystal lattice.\ Importantly, a description of the physics of such disorder entails methods from statistical physics where no attempt is done to account for each individual imperfection, but rather light transport is described by universal (macroscopic) parameters of the system such as the mean-free path~\cite{Sheng,Rossum}.\ It turns out that disorder is particularly critical at certain frequencies in the dispersion diagram, e.g., near the cut-off frequency of the waveguide mode (cf. dashed line in Fig.~\ref{2}\textbf{c}) or at the edges of the bandgap since the associated highly-dispersive slow-light optical modes become very sensitive to disorder.\ As such, photonic crystals constitute an interesting example of a system possessing both order an disorder.\ While the underlying ordered lattice gives rise to the tailored dispersion relation with, e.g., band gaps and slow light, the presence of disorder perturbs the modes leading to random multiple scattering and localization effects.\ The underlying ordered photonic-crystal lattice is beneficial for inducing Anderson localization since strong dispersion of light may aid in crossing the boundary to the strongly-localized regime, which was already proposed in a pioneering work on photonic crystals~\cite{John}.\ In photonic-crystal waveguides, this close relation between the group velocity and the sensitivity to imperfections was realized by S. Hughes and coworkers~\cite{Hughes}.

\begin{figure}[b!] \includegraphics[width=\columnwidth]{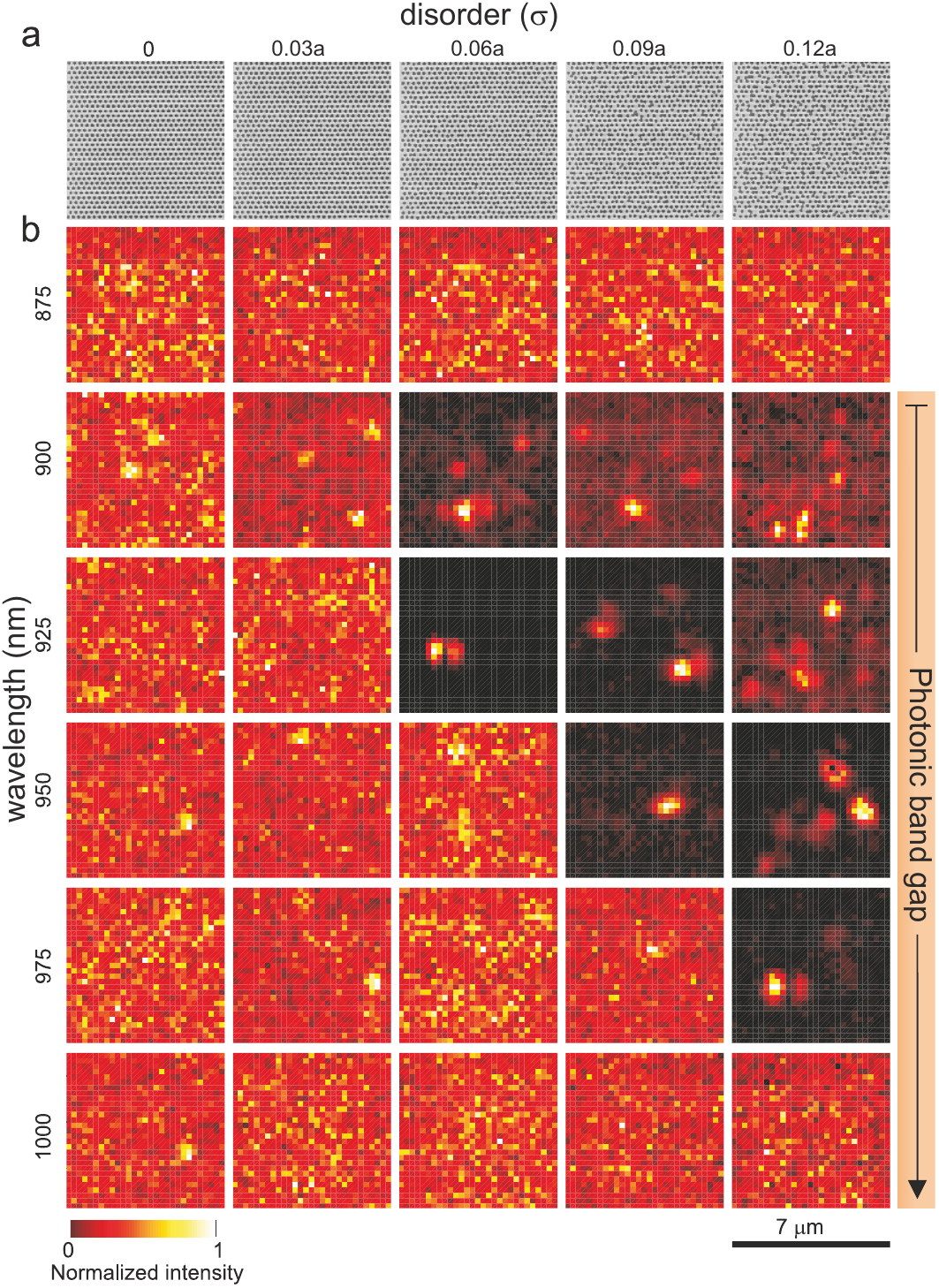} \caption{ \label{3}
\textbf{a}.\ Scanning electron micrographs (top view) of photonic-crystal membranes with $a=280\,\text{nm}$, $r=0.274a$, perturbed by an amount of disorder varying between $\sigma = 0$ and $\sigma = 0.12a$.
\textbf{b}.\ High-power photoluminescence spectra are collected while scanning the samples.\ At each position, the spectrum is normalized and wavelength-binned with a bin size of $1\,\text{nm}$.\ A set of these wavelength-binned photoluminescence scans are plotted from $875\,\text{nm}$ to $1000\,\text{nm}$ every $5\,\text{nm}$.\ The orange arrow outlines the spectral position of the photonic bandgap.\ From Ref.~\cite{C0}.} \end{figure}

To establish their performance limitations, it is an essential task to quantify systematically the robustness of photonic-crystal nanostructures towards imperfections~\cite{Gerace}.\ Figures~\ref{2}\textbf{d} and~\ref{2}\textbf{e} show examples of structures where an intentional amount of disorder is introduced in the hole positions in accordance to a Gaussian probability distribution with a standard deviation, $\sigma = \sqrt{\langle \Delta \text{\textbf{r}}^2\rangle}$, where $\Delta \text{\textbf{r}}$ is the  random hole displacement.\ Here, $\sigma=0$ corresponds to no intentional disorder, but intrinsic fabrication disorder in the samples is nonetheless unavoidable.\ The photonic modes in the disordered structures are conveniently probed by using the internally embedded ensemble of emitters (InAs quantum dots) to excite the modes.\ As the size of the InAs self-assembled quantum dots is about ten nanometer and with a refractive index close to that of the matrix (GaAs), their role on the scattering properties of the structure can be considered negligible.\ This is particularly true in the situation where the quantum dots are strongly pumped optically into saturation, which is usually the experimental condition under which the modes of the nanostructure are probed.\ For such experiments a relatively large density of quantum emitters is beneficial, since the inhomogeneously broadened photoluminescence (PL) due to the quantum dot size distribution can be exploited as an embedded broadband light source.\ By scanning across the sample and collecting the PL spectra, the statistics of the modes in the structures can be probed.\ Since the specific quantum dots used in the present example are operated optimally at low temperatures, the experiment was carried out at $\text{T} = 10\,\text{K}$ in a Helium flow cryostat and the sample position was controlled with stages with a resolution of $0.3\,\micro\text{m}$.\ The PL spectrum was sent through a spectrometer and detected either with a CCD camera for spectral analysis or with an avalanche photo-diode for the dynamics (see details in Refs.~\cite{Luca,C0}).\ Figure~\ref{3} plots the spatially-resolved PL intensity collected at different wavelengths from photonic crystals perturbed by an increasing amount of disorder from $\sigma = 0$ to $\sigma = 0.12a$, with $a=280\,\text{nm}$ being the lattice constant of the photonic crystal.\ The effect of disorder is apparent: while the PL fluctuates weakly in slightly perturbed lattices ($\sigma \leq 0.03a$), the presence of very bright peaks is observed for $\sigma \geq 0.06a$.\ This is the manifestation of two-dimensional Anderson localization of light observed near the edge of the photonic band gap.\ Localization occurs within a certain wavelength range close to the photonic band gap, corresponding to the so-called optical Lifshitz tail~\cite{Lifshitz}, which has been experimentally studied also in one-dimensional photonic-crystal waveguide systems by scanning near-field optical microscopy~\cite{Huisman} or single quantum-emitter spectroscopy~\cite{TyrrestrupAPL}.

\begin{figure}[t] \includegraphics[width=\columnwidth]{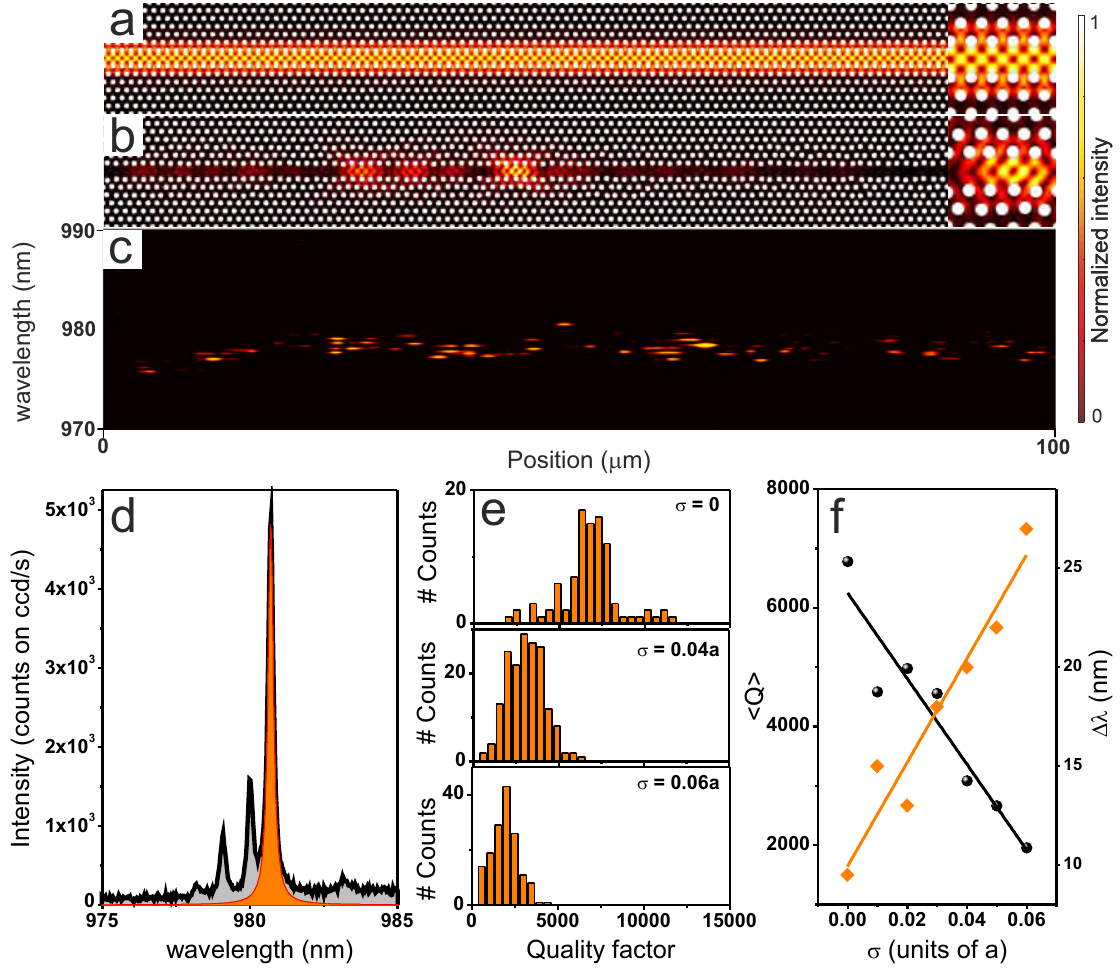} \caption{ \label{4}
Finite-difference time-domain calculation of the electromagnetic-field intensity in an ideal (\textbf{a}) and in a disordered (\textbf{b}) photonic-crystal waveguide with a lattice constant $a=260\,\text{nm}$ and a hole radius 0.29a, calculated at the cutoff frequency of waveguide mode.\ The disordered waveguide is perturbed by $\sigma=0.04a$.\ \textbf{c}.\ Experimentally recorded high-power photoluminescence spectra collected while scanning a microscope objective along a photonic-crystal waveguide with only intrinsic disorder ($\sigma = 0$).\ \textbf{d}.\ Example of a photoluminescence spectrum collected at a fixed position, where the red area is a Lorentzian fit to one cavity resonance.\ \textbf{e}.\ Experimental Q-factor distributions of the modes measured in photonic-crystal waveguides (histograms) perturbed by different amount of disorder.\ \textbf{f}.\ The ensemble-averaged Q factor and the spectral range where the Anderson-localized modes are observed (i.e. the width of the Lifshitz tail)
as a function of $\sigma$.\ From Ref.~\cite{Smolka}.} \end{figure}

As opposed to 2D photonic-crystals slabs, which are relatively robust against disorder, planar photonic-crystal waveguides require just a small amount of residual disorder to obtain Anderson localization in the wavelength range where light propagation is slow~\cite{Luca}.\ Figure~\ref{4}\textbf{a} displays the electric field intensity in an ideal photonic-crystal waveguide calculated at the cutoff frequency of the guided mode by the numerical method described in Ref.~\cite{mpb}.\ Near this frequency, light is slowed down, i.e., the group velocity ideally vanishes at the cutoff point~\cite{Baba}, an effect that has proven well suited for enhancing light-matter interaction~\cite{Toke,Marta}.\ The effect of disorder on the waveguide mode near the cutoff is illustrated in the calculation presented in Fig.~\ref{4}\textbf{b}.\ The periodic Bloch mode is observed to split up due the effects of random multiple scattering, whereby a collection of strongly localized cavities appear.\ In an experiment they may be excited by the internal emitters (e.g., quantum dots) and Fig.~\ref{4}\textbf{c} shows an example of a recorded emission spectrum.\ More than hundred Anderson-localized modes were measured by scanning along the waveguide and the data are analyzed, e.g., by extracting the cavity quality factor $\text{Q}=\omega/\Delta\omega$, which measures the ratio between the cavity frequency and linewidth.\ $\text{Q}$ is a relevant parameter for cavity quantum electrodynamics (QED) since it determines, jointly with the mode volume $V$, the Purcell enhancement of an emitter on perfect spatial and spectral resonance with the cavity as~\cite{Purcell}:

\begin{equation*}
 \text{F}_\text{p}  = \frac{3}{4 \pi^2}  \left( \frac{\lambda}{n} \right)^3 \frac{ Q }{ V },
\end{equation*}

where $\lambda$ is the wavelength in vacuum and $n$ the refractive index.

A broad distribution of $\text{Q}$ factors ranging from $\text{Q}=200$ to $\text{Q}=13.000$ was recorded experimentally, where the distribution is found to strongly depend on the amount of disorder introduced to the sample, cf. Fig.~\ref{4}\textbf{e}.\ Interestingly, the highest Q-factors were observed in waveguides where no extrinsic disorder was introduced ($\sigma = 0$), i.e. intrinsic disorder due to inherent fabrication imperfections suffices in creating high-Q random cavities albeit occurring in a narrow bandwidth in the slow-light regime, which is quantified as the width of the Lifshitz tail.\ Two mechanisms determine $\text{Q}$ in the waveguides: the in-plane loss from light leaking out of the end of the waveguide due to the finite length and out-of-plane leakage due to scattering out of the membrane structure~\cite{Smolka}.\ Figure~\ref{4}\textbf{f} shows that the ensemble-averaged cavity Q-factor decreases with the amount of disorder $\sigma$.\ In contrast, the width of the Lifshitz tail, $\triangle\lambda_L$, increasea with disorder, i.e. the spectral range in which Anderson localization is observed to increase with disorder.\ By using the experimentally recorded $\triangle\lambda_L$ as a measure of the amount of disorder, it is possible to quantify the amount of intrinsic disorder, i.e., the disorder due to the fabrication limitations.\ In photonic crystals, intrinsic disorder is a combination of different types of imperfections such as fluctuations in the hole positions, radii, and shape.\ The amount of disorder in state-of-the-art photonic-crystal waveguides has been  quantified to correspond to a hole displacement of $\sigma \approx 0.005a$~\cite{quantifying}, which is a valuable quantitative measure of the influence of statistical fluctuations on the propagation of light in the waveguide.

\section{P\lowercase{robing the statistical properties of} A\lowercase{nderson localization with internal emitters.}}

\begin{figure}[b] \includegraphics[width=\columnwidth]{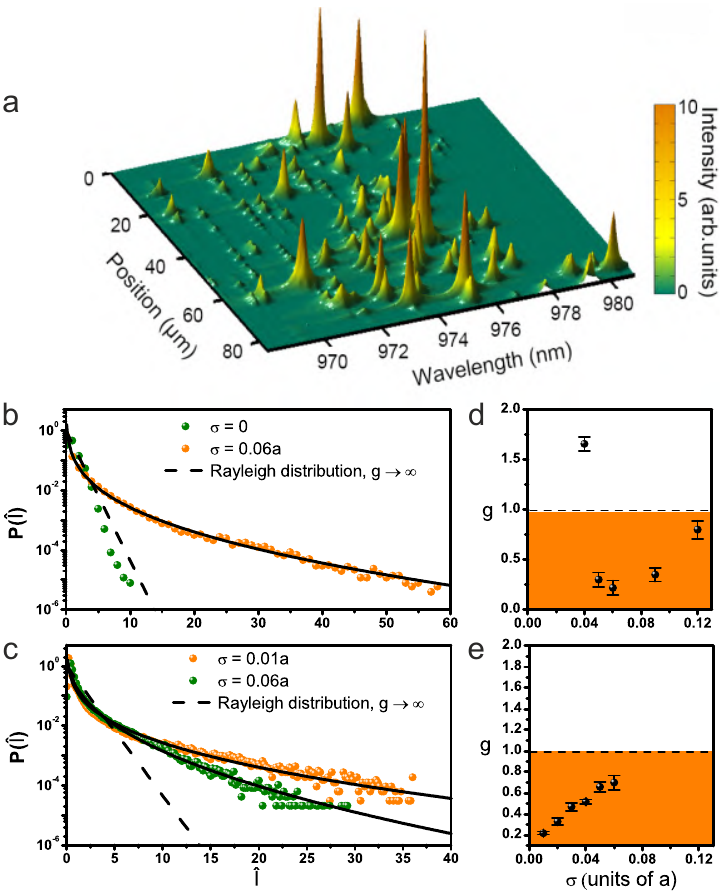} \caption{ \label{5}
\textbf{a}.\ High-power photoluminescence spectra collected while scanning an excitation/collection objective along a photonic-crystal waveguide with an amount of disorder of $\sigma = 0.03a$.\ From Ref.~\cite{Luca}.\ Normalized intensity distribution recorded from the spectra for the case of a $7\,\micro\text{m}\,\times\,7\,\micro\text{m}$ 2D photonic crystal (\textbf{b}) and a $100\,\micro\text{m}$-long photonic-crystal waveguide (\textbf{c}) and varying amounts of disorder.\ A pronounced deviation from Rayleigh statistics (dashed line) is observed for disordered samples.\ The solid lines represent the best fit to the theory of Ref.~\citep{Rossum} from which the dimensionless conductance, $\textit{g}$ can be obtained.\ \textbf{d} and \textbf{e} show the recorded $\textit{g}$ versus amount of disorder in a photonic crystal and a photonic-crystal waveguide, respectively.\ From Ref.~\cite{C0} and Ref.~\cite{Smolka}.} \end{figure}

Anderson localization is in essence a statistical phenomenon and, therefore, it is important to extract reliably the statistical properties of light transport in this regime.\ The standard approach is to study reflection and transmission through a disordered medium, which is of limited use for probing strongly localized modes that are hardly accessible with external sources.\ Thus, in such transmission experiments, only confined modes with sizeable amplitudes at the sample surface can be excited, thus imposing a limitation on a detailed statistical analysis.\ The incorporation of light emitters homogenously distributed inside the samples turns out to be a major advantage, since they may efficiently excite both propagating and localized modes.\ As a consequence, the statistics of disordered optical modes can be recorded by monitoring the intensity fluctuations and spectral components of the light emitted by the internal light sources, e.g., an ensemble of quantum dots~\cite{Smolka}.

\begin{figure*}[t!] \includegraphics[width=\textwidth]{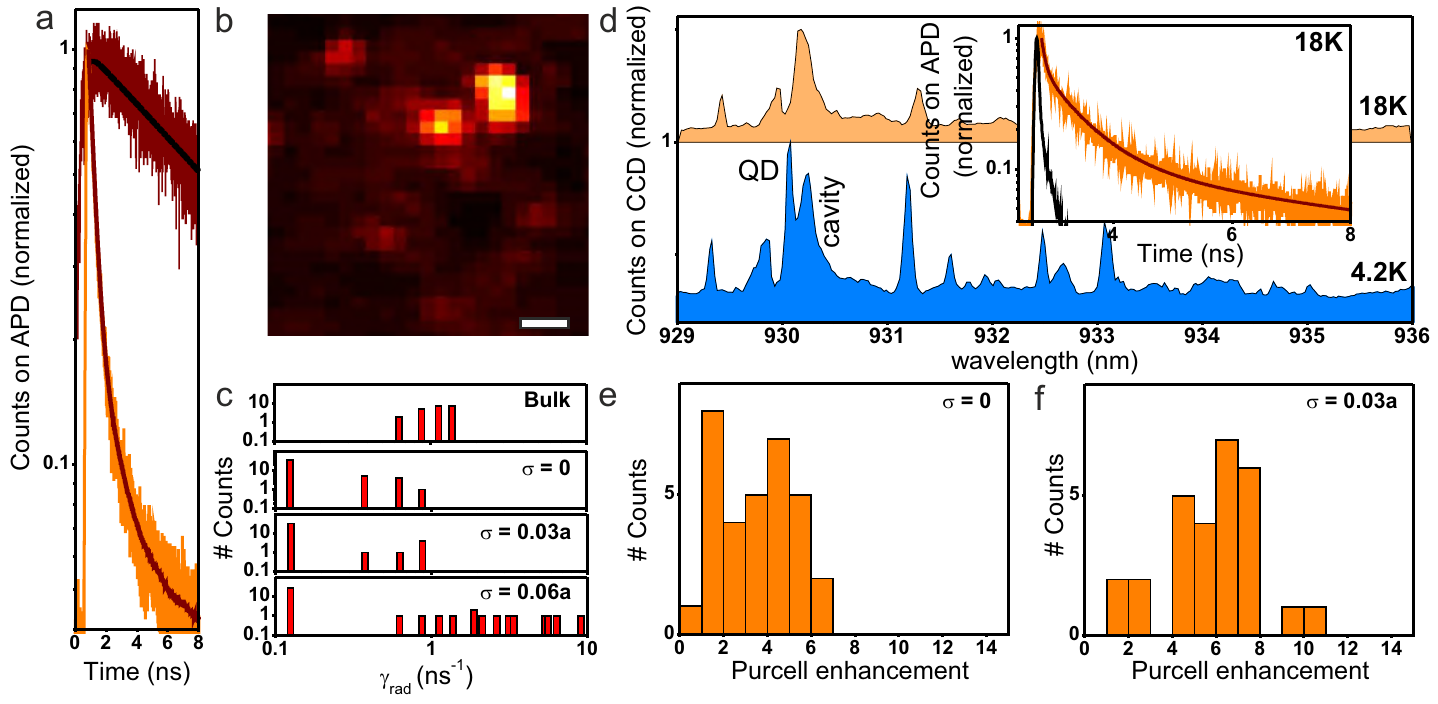} \caption{ \label{6}
\textbf{a}.\ Decay curves for two quantum dots placed at different positions in a photonic crystal perturbed by $\sigma = 0.06a$ and the fits to bi-exponential models (solid lines).\ \textbf{b}. High-power PL spectra collected from the photonic crystal revealing Anderson-localized modes. \textbf{c}.\ Radiative decay rate ($\gamma_\text{rad}$) distributions measured at a wavelength of  $\lambda = 910\pm3\,\text{nm}$ in the bulk membrane and in photonic crystals affected by different amounts of disorder.\ \textbf{d}.\ PL spectrum from a photonic-crystal waveguide at two different temperatures at low-power excitation, where both single quantum dot (QD) lines and Anderson-localized cavities can be identified.\ The inset plots the decay curve recorded from the quantum dot on resonance with the cavity at $\lambda = 917\,\text{nm}$ corresponding to T = 18K, where the black curve is the instrument-response function of the detector.\ \textbf{e} and \textbf{f}. Histograms displaying the Purcell enhancement measured on Anderson-localized modes of a photonic-crystal waveguide for $\sigma=0$  and $\sigma=0.03a$, respectively.} \end{figure*}

Recording the intensity of an external light source to characterize a random medium in a transmission experiment has an additional drawback.\ In such a configuration, discriminating between Anderson localization and simple absorption is subtle because both give rise to similar effects on the averaged transmission~\cite{red}.\ A statistical approach was proposed to overcome this issue and was implemented for the case of microwaves~\cite{Chabanov}, where the statistical fluctuations of the multiply scattered transmitted light was recorded and analyzed.\ Having internal light sources inside the disordered structure allows to generalize this approach by constituting a convenient method of exciting the modes of a random medium and therefore extract the statistics.\ Figure~\ref{5}\textbf{a} shows an example of such measurements for the case of a disordered photonic-crystal waveguide where the internal modes in the Anderson-localized regime are excited by the emitters and recorded in the PL spectra.\ From such data, the PL intensity distribution $\text{P}(\widehat{ \text{I}} \equiv \text{I}/\langle \text{I} \rangle)$ can be extracted, where $\langle \text{I} \rangle$ is the average PL intensity.\ For a moderate effect of disorder, random multiple scattering is negligible and the PL intensity is described by Rayleigh statistics~\cite{Sheng}.\ For increased disorder, however, $\text{P}(\widehat{\text{I}})$ evolves into a heavy-tailed distribution, revealing the presence of few but bright peaks in a low-intensity background.\ Examples of experimentally recorded distributions are plotted in Fig.~\ref{5} for two-dimensional photonic crystals (\textbf{b}) and one-dimensional photonic-crystal waveguides (\textbf{c}), respectively and modeled by the associated theory \cite{Rossum,Tigelen}.\ This analysis is reliably confirming effects of multiple scattering  since absorption cannot give rise to such a heavy-tailed intensity distribution.\

A brief account of the theory applied to the experimental data comes as follows; in the context of electronic transport, the scaling theory~\cite{g} proposes a single parameter, the Thouless or dimensionless conductance $\textit{g}$, to describe the conductor-insulator phase transition induced by disorder.\ This theory can be extended to the case of light~\cite{Sheng,Rossum}. The dimensionless conductance $\textit{g}$ can be defined as the total transmittance, i.e., the sum over all the transmission coefficients connecting all input-output modes.\ This parameter is particularly interesting since it governs all the statistical aspects of light transport in a random medium~\cite{Chabanov}, and sets the boundary $(\textit{g} \leq 1)$ for Anderson localization in absence of absorption\ $\textit{g}$ can be determined by fitting the experimental PL intensity probability distribution with the theory developed van Rossum and Neuwenhuizen in Ref.~\citep{Rossum}
\begin{equation}\label{intensity_distributoion}
\text{P}(\widehat{\text{I}})= \int_{- i \infty}^{i \infty} \! \frac{\mathrm{d} x}{\pi i} \text{K}_0 (2 \sqrt{-\widehat{\text{I}}x}) e^{-\Phi_{\text{con}}(x)},
\end{equation}
 which is valid in the regime of perturbative scattering and in absence of absorption. Here $\text{K}_0$ is a modified Bessel function of second kind and $\Phi_{\text{con}}(x)$ is obtained by assuming plane-wave incidence to be:
\begin{equation}\label{intensity_distributoion}
\Phi_{\text{con}}(x)= \textit{g} \ln^2(\sqrt{1 + \frac{x}{\textit{g}}} +\sqrt{\frac{x}{\textit{g}}})
\end{equation}

The variation of $\textit{g}$ with disorder is plotted in Fig.~\ref{5} for photonic crystals (\textbf{d}) and photonic-crystal waveguides (\textbf{e}), respectively.\ The wave transport in 2D photonic-crystal slabs is less sensitive to disorder than in photonic-crystal waveguides and displays a different functional dependence in the range of disorder covered in these experiments.\ When imperfections are introduced in a lattice, $\textit{g}$ naturally decreases as random multiple scattering increases.\ In this case, the effect of disorder may be enhanced by slow light due to the presence of the underlying lattice~\cite{John}.\ However, this is only valid when imperfections are a perturbation to the lattice.\ When the amount of imperfections is sufficiently large, the enhancement due to slow-light effects disappears and multiple scattering effects are reduced.\ Following this interpretation, a minimum should be found in $\textit{g}$ corresponding to an optimum amount of disorder, i.e., sufficiently high to provoke strong scattering while preserving the lattice.\ This optimum reflects an interesting balance between order and disorder: the lattice gives rise to dispersion and slow light that in turn is more sensitive to disorder.\ As shown in Fig.~\ref{5}\textbf{d} and \textbf{e}, this optimum amount of disorder corresponds to $\sigma \sim 0.06a$ in the photonic crystals and it is less than the intrinsic fabrication imperfection in the photonic-crystal waveguides.

\section{C\lowercase{avity quantum-electrodynamic experiments and lasing.}}

The appearance of localized modes in the photonic structures may strongly affect the emission dynamics of embedded light emitters.\ This is the basis for cavity quantum-electrodynamics experiments with disordered cavities.\ The radiative decay rate of a dipole emitter is directly related to the projected LDOS at the position of the emitter, $\textbf{r}$, i.e., $\gamma_\text{rad}(\textbf{r},\omega)\propto|\textbf{p}|^2\text{LDOS}(\textbf{r},\omega)$, where $\textbf{p}$ is the emitter dipole moment.\ This is the basis for the construction of low-threshold lasers~\cite{Noda}, deterministic single-photon sources~\cite{Marta}, and light-matter quantum entanglement~\cite{Strong}.\ Quantum dots embedded in the samples may be exploited as probes of the LDOS and its spatial fluctuations.\ In this case the quantum dots are pumped at low excitation power with short optical pulses whereby the dynamics of single quantum dot lines can be probed.\ The emission curves for neutral excitons in InAs quantum dots pumped by non-resonant excitation are bi-exponential as a consequence of the fine structure of the lowest exciton state~\cite{reviewMod}, which allows to extract the radiative decay rate of single emitters at a given wavelength and position~\cite{Qin1}.\ It is worth emphasizing that intrinsic non-radiative processes in the quantum dots, corresponding to an imperfect internal quantum efficiency, may explicitly be accounted for in the decay curve analysis.\ Consequently, the intrinsic radiative decay rate of the quantum dots can be extracted and the LDOS sensitively mapped~\cite{Qin1}.\ Figure~\ref{6}\textbf{a} plots the time-resolved PL at a wavelength of $\lambda = 910\pm3\,\text{nm}$ for two quantum dots  at different positions in a photonic crystal perturbed by $\sigma = 0.06a$.\ By applying a high excitation power to the same sample, the localized modes in the photonic crystal are revealed, cf. Fig.~\ref{6}\textbf{b}.\ Strongly localized modes are found, and the rapid decay observed for one of the quantum dots in Fig.~\ref{6}\textbf{a} is due to the coupling to one of the Anderson-localized modes of the structure.\ The statistics of the recorded radiative decay rates for single quantum dots is reproduced in Fig.~\ref{6}\textbf{c}, which displays a significant broadening of the radiative emission decay rate distributions in the localization regime.\ It clearly illustrates that the LDOS can be strongly modified in disordered photonic nanostructures enabling both large Purcell factors as well as strong suppression of spontaneous emission. A full mapping of the spatial and spectral variation of the LDOS in disordered photonic crystals and photonic-crystal waveguides remains a future challenge, connecting to the fundamental open question of how large LDOS modifications may be obtained in 2D disordered nanostructures.

\begin{figure}[t!] \includegraphics[width=\columnwidth]{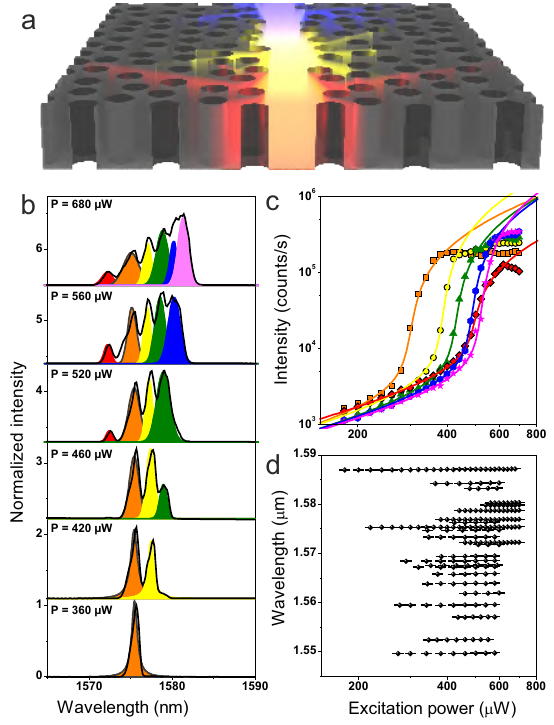} \caption{ \label{7}
\textbf{a}.\ Illustration of a random laser in a disordered photonic-crystal waveguide formed from Anderson-localized modes, where the different colors indicate multimode lasing operation.\ \textbf{b}.\ PL spectra versus excitation power in multimode operation. Here the different colored curves are fits with Lorentzian functions representing different lasing modes.\ The spectra taken at different powers are offset for clarity. \textbf{c}.\ Output intensity versus excitation power corresponding to the different lasing modes in \textbf{b} where the solid lines represent fits to semiconductor laser rate equations.\ \textbf{d}.\ Central wavelength obtained from the Lorentzian fits of all the measured modes versus excitation power.\ From Ref.~\cite{Jin}.} \end{figure}

A direct measure of the Purcell enhancement in Anderson-localized modes requires control over the detuning of the quantum dot relative to the cavity, which can be implemented by temperature tuning~\cite{kiraz}.\ Fig.~\ref{6}\textbf{d} shows  two PL spectra at different temperatures for a photonic-crystal waveguide affected only by intrinsic disorder.\ Emission lines corresponding to either single quantum dots or Anderson-localized modes  can be identified from their different temperature dependencies. Figure~\ref{6}\textbf{d} shows spectra and decay curves for a single quantum dot tuned into resonance with a localized mode displaying a Purcell factor as high as 23.8.\ This is identified as being close to the onset of strong coupling~\cite{Strong_Carminati,Strong_Tyrrestrup,Wong}.\ Unfortunately, such detailed tuning experiments are rather cumbersome and therefore extracting the full statistics of the Purcell enhancement by this method is not practical.\ An alternative approach records the time-resolved emission directly from the Anderson-localized cavities at a fixed temperature~\cite{Javadi}.\ In such measurements, contributions from several quantum dots coupled to the cavity modes lead to a multi-exponential decay.\ By extracting only the fastest component of the decay curves corresponds to investigating the contributions from the dominating quantum dot.\ The Purcell factor is derived by referencing these data to the decay rate of quantum dots measured in the GaAs homogeneous environment, which is 1.1 ns\textsuperscript{-1}.\ Figures~\ref{6}\textbf{e} and \ref{6}\textbf{f} show Purcell factor distributions for different amounts of disorder, where the data notably constitute a lower bound since the detuning between the quantum dots and the cavity modes is not controlled.\ A detailed theory for the Purcell factor distributions remains an open question, which would require a detailed LDOS description in disordered photonic-crystal waveguides.

\begin{figure}[t!] \includegraphics[width=\columnwidth]{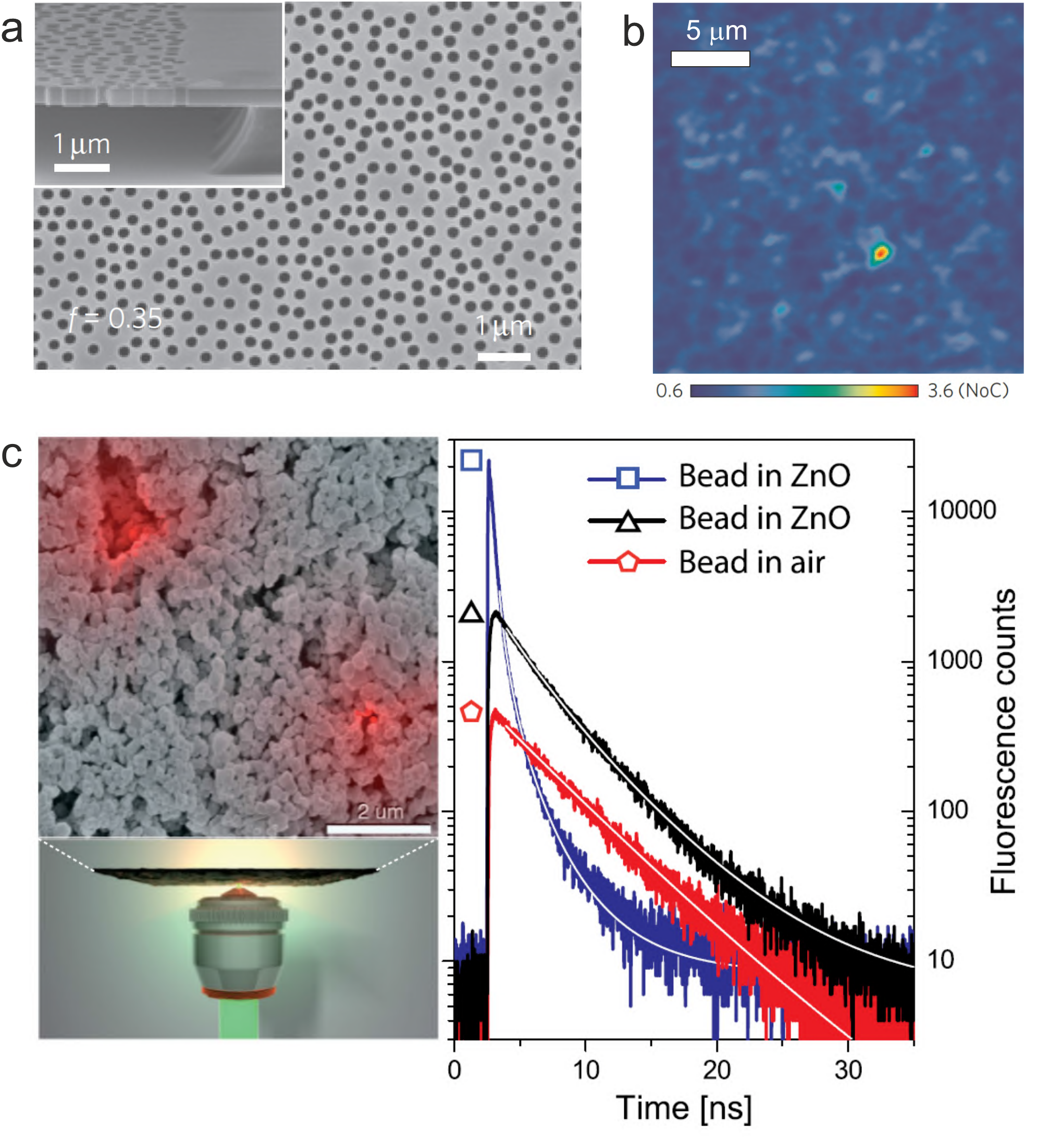} \caption{ \label{8}
\textbf{a}.\ Scanning electron micrograph (top view) of amorphous disordered structures made by etching holes randomly in GaAs.\ \textbf{b}.\ Near-field photoluminescence images of the sample shown in \textbf{a} at $\lambda = 1300\,\text{nm}$.\ From Ref.~\cite{Riboli}.\ \textbf{c}.\ The left panel is a schematic of the experiment used to obtain scanning electron microscope images of disordered ZnO.\ The right panel plots the decay curve of different ensembles of emitters in air and in a ZnO matrix.\ From Ref.~\cite{Ricca}.} \end{figure}

Finally, introducing optical gain into disordered structures leads to new opportunities by investigating the formation of random lasing ~\cite{Frank}.\ To this end, disordered photonic-crystal waveguides constitute a suitable platform~\cite{APL} where random lasing in the Anderson-localized regime has been achieved~\cite{Jin}.\ Figure~\ref{7}\textbf{a} illustrates Anderson-localized random lasing in a disordered photonic-crystal waveguide.\ In such experiments, it is favorable to exploit quantum wells as active medium as opposed to quantum dots in order to achieve a sufficiently large gain.\ Figures~\ref{7}\textbf{b-d} show results of such an experiment carried out on InGaAsP quantum wells embedded in InP membranes and operated at room temperature.\ The quantum well is optically pumped and by increasing excitation power several localized modes start to lase in a sequential manner, i.e. the different laser peaks grow one by one with increasing excitation power each displaying a characteristic input-output lasing curve.\ Figure~\ref{7}\textbf{d} plots the multi-mode lasing wavelengths, which are remarkably stable with increasing power even for modes that are spectrally and spatially close to each other.\ Such stable multimode operation is a feature of operating in the Anderson-localized regime: since the modes are exponentially confined (on average) the corresponding coupling between different lasing is suppressed due to their weak spatial overlap~\cite{Stano2012}.\ On the contrary, random lasing modes found in diffusive systems are largely extended and therefore overlap spatially implying that different modes deplete the same gain region.\ This is found to lead to very strong mode competition and eventually chaotic lasing behavior~\cite{Cahotic_lasing1,Cahotic_lasing2}.

\section{E\lowercase{mitters in amorphous structures.}}

The use of light emitters to probe the properties of disordered dielectric structures is not limited to perturbed lattices.\ Figure~\ref{8}\textbf{a} and \textbf{b} illustrate the use of two-dimensional amorphous nanostructures~\cite{Riboli} instead of the photonic crystals previously discussed.\ In these GaAs nanostructures, which also include embedded InAs quantum dots, the etched holes are randomly distributed with a controlled density and a minimum distance between them, as shown in Fig.~\ref{8}\textbf{a}.\ The PL from the internal quantum dots is monitored by using a near-field scanning optical microscope as shown in Fig.~\ref{8}\textbf{b}, revealing the presence of strongly localized cavities with a high degree of spatial confinement.\ Near-field optical microscopy has been proven as a powerful method of characterizing and visualizing the highly complex spatial mode distributions found in disordered photonic nanostructures~\cite{Huisman,Intonti,Spasenovic}.\ Since a minimum distance between the scatterers is set by the hole size, such a structure has both short- and long-range structural correlations, which relate to the propagation of light~\cite{Matteo} and can be used to tailor the LDOS of the system~\cite{correlations}.\ Also in the diffusive regime, the emission dynamics of emitters embedded in complex photonic networks is significantly affected.\ One experiment applied fluorescent organic dye molecules embedded in a matrix of ZnO powder~\cite{Ricca} where the scattering properties of the material modify the emission dynamics of the emitters, as shown in Figure~\ref{8}\textbf{c}.\ Modified spontaneous emission dynamics has been reported in strongly diffusive materials such as ZnO~\cite{Ricca,Vos} or $\text{TiO}_{2}$~\cite{Sandoghdar}, which gives rise to a broadening of the emission rate distribution induced by fluctuations of the LDOS in the dielectric matrix.

\section{C\lowercase{onclusions and outlook.}}

In summary, we have presented a brief Review of some recent results on the physics of quantum light emitters in disordered photonic nanostructures.\ The role of unavoidable fabrication disorder is essential to comprehend in order to exploit photonic nanostructures in, e.g., quantum optics~\cite{reviewMod}.\ The onset of disorder occurs on a characteristic length scale, the mean-free path, which is the average propagation distance in between scattering events. Statistical measurements, as the ones presented here, grant access to such universal transport phenomena.\ To circumvent disorder effects, it is therefore required to make nanostructures that are smaller than the mean-free path, which turns out to be a viable approach in state-of-the-art photonic nanostructures, as only few unit cells are sufficient to build-up photonic-crystal effects.\ Remarkably, an alternative and less intuitive approach to nanophotonics and quantum optics may be taken that exploits disorder as a mean to enhance the interaction between light and matter.\ Some recent results on the applications of quantum emitters in disordered photonic nanostructures have been highlighted in the present manuscript.\ A benefit of this approach is that the embedded emitters are very convenient probes of the complex statistical properties of a disordered medium.\ Current research topics include the study of mesoscopic quantum correlations in disordered media~\cite{Lodahl2005} where  non-universal correlations may occur when incorporating light emitters~\cite{C0,Saphiro,imaging,CO_Carminati,correlations,Ricca} that have no counterpart in mesoscopic electronics.\ The approach of exploiting disorder to induce functionalities - \emph{disordered photonics}~\cite{Wiersma2} - emerges with potential applications in, e.g., near-field imaging~\cite{imaging}, lasing~\cite{Wiersma1,Frank,Cahotic_lasing1,Cahotic_lasing1}, cavity quantum-electrodynamics~\cite{Luca}, or energy harvesting~\cite{Vynk,Peretti,Noda2}.

\begin{figure}[t!] \includegraphics[width=\columnwidth]{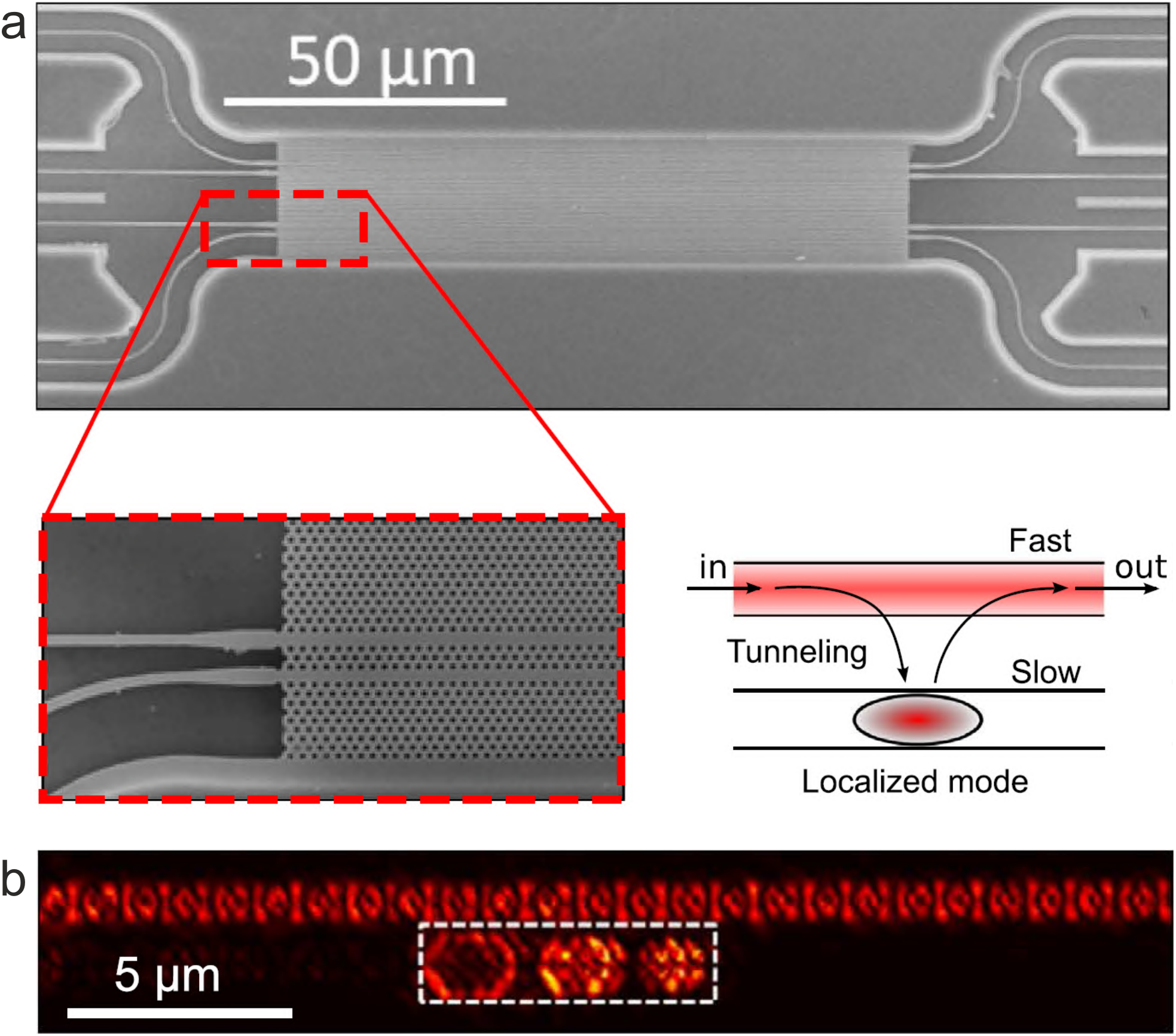} \caption{ \label{9}
\textbf{a}.\ Scanning electron microscope image of two side-coupled photonic-crystal waveguides and of the access ridge waveguides used to inject light.\ Light is injected in the upper waveguide, which operates in a regime where light propagation is fast, and couples evanescently to localized modes in the lower waveguide, operating in the slow-light regime, which is illustrated in the sketch.\ \textbf{b}.\ High-resolution near-field images of the photonic-crystal waveguide pair recorded at $\lambda = 1489.61\,\text{nm}$, which shows an extended mode coupled to a localized mode.\ From Ref.~\cite{Lalanne2}.} \end{figure}

It is an interesting question to consider to what extent disordered photonics offers solutions to future challenges in quantum photonics and nanophotonics.\ One major issue is to scale-up simple functionalities to more advanced architectures.\ For instance, in quantum photonics a network of coupled cavities would be required to scale-up cavity quantum-electrodynamics.\ Disordered photonics offers a potential approach to that by exploiting statistically-occurring coupled Anderson-localized modes, the so-called necklace states~\cite{Bertolotti}.\ By controlling the universal statistical parameters determining light transport, the likelihood of creating a certain necklace state could be enhanced.\ Such necklace states have been observed in amorphous dielectric nansotructures~\cite{Riboli} and have also explored in perturbed lattices~\cite{Toninelli}, and could provide an interesting approach to study the physics of arrays of coupled cavities.\ One challenge for disordered cavities would be to couple light in and out with a high efficiency, which would be essential for low-loss devices.\ However, this could conveniently be implemented on-chip by positioning next to an Anderson-localizing waveguide a second waveguide, which is tailored to be in a regime where light propagation is fast and therefore localization would not occur.\ Nonetheless the coupling between the two waveguides can be precisely controlled by exploiting evanescent coupling, and an illustrative implementation of this scheme is shown in Fig.~\ref{9}\textbf{a}.\ In this case light propagating in the access waveguide was evanescently coupled to strongly localized modes in the slow-light waveguide, which was observed in the near-field measurement reproduced in Fig.~\ref{9}\textbf{b}.\ This configuration could also enable a highly efficient and stable multi-color random laser with directional output.

\textbf{Acknowledgements}

It is a pleasure to acknowledge the very fruitful collaborations we have had over the years on disorder in nanophotonic structures with many talented researchers including H. Thyrrestrup, A. Javadi, S. Smolka, L. Sapienza, S. Stobbe, J. Liu, S.~R. Huisman, W.~L. Vos, P.~W.~H. Pinkse, N. Mann, and S. Hughes.\ We would also like to thank Kevin Vynk for his constructive comments on the manuscript.
We gratefully acknowledge financial support from the Villum Foundation, the European Research Council (ERC Consolidator Grant ''ALLQUANTUM''), and the Danish Council for Independent Research.



\begin{thebibliography}{99}


\bibitem{Purcell}
E.~M. Purcell.
\emph{Spontaneous emission probabilities at radio frequencies}.
Phys. Rev. \textbf{69}, 681 (1946).

\bibitem{Vahala}
K.~J. Vahala.
\emph{Optical microcavities}.
Nature \textbf{424}, 839 (2003).

\bibitem{Joannopoulos}
J.~D. Joannopoulos, S.~G. Johnson, Jo.~N. Winn, and R.~D. Meade.
\emph{Photonic crystals: molding the flow of light. Princeton}.
 N.J., Princeton University Press.

\bibitem{Vucovick}
J.~L. O'Brien, A. Furusawa, and J. Vu\v{c}kovi\'{c}.
\emph{Photonic quantum technologies}.
Nat. Phot. \textbf{3}, 687 (2009).

\bibitem{Akahane}
Y. Akahane, T. Asano, B. Song, and S. Noda.
\emph{High-Q photonic crystal nanocavity in a two-dimensional photonic crystal}.
Nature \textbf{425}, 944 (2003).

\bibitem{Savona_optimization}
M. Minkov, and V. Savona.
\emph{Automated optimization of photonic crystal slab cavities}.
Scientific Reports \textbf{4}, 5124 (2014).

\bibitem{Luca}
L. Sapienza, H. Thyrrestrup, S. Stobbe, P.~D. Garc\'{i}a, S. Smolka, and P. Lodahl.
\emph{Cavity quantum electrodynamics with Anderson-localized modes}.
Science \textbf{327}, 1352 (2010).

\bibitem{ALlight}
M. Segev, Y. Silberberg, and D.~N. Cristodoulides.
\emph{Anderson localization of light}.
Nat. Phot. \textbf{7}, 197 (2013).

\bibitem{Lawandy}
N.~M. Lawandy, R.~M. Balachandran, A.~S.~L. Gomes, and E. Sauvain.
\emph{Laser action in strongly scattering media}.
Nature \textbf{368}, 436 (1994).

\bibitem{Wiersma1}
D.~S. Wiersma.
\textit{The physics and applications of random lasers}.
Nat. Phys. \textbf{4}, 359 (2008).

\bibitem{Wiersma2}
D.~S. Wiersma.
\textit{Disordered photonics}.
Nat. Phot. \textbf{7}, 188 (2013).

\bibitem{Koenderink}
A.~F. Koenderink, and W.~L. Vos.
\emph{Light exiting from real photonic band gap crystals is diffuse and strongly directional}.
Phys. Rev. Lett. \textbf{91}, 213902 (2003).

\bibitem{PCdisorder}
P.~D. Garc\'{i}a, R. Sapienza, C. Toninelli, C.~L\'{o}pez, and D.~S. Wiersma.
\emph{Photonic crystals with controlled disorder}.
Phys. Rev. A \textbf{84}, 023813 (2011).

\bibitem{Lalanne1}
S. Mazoyer, J.~P. Hugonin, and P. Lalanne.
\emph{Disorder-induced multiple scattering in photonic-crystal waveguides}.
Phys. Rev. Lett. \textbf{103}, 063903 (2009).

\bibitem{Vollmer}
J. Topolancik, B. Ilic, and F. Vollmer.
\emph{Experimental observation of strong photon localization in disordered photonic crystal waveguides}.
Phys. Rev. Lett. \textbf{99}, 253901 (2007).

\bibitem{Hughes}
S. Hughes, L. Ramunno, J.~F. Young, and J.~E. Sipe.
\emph{Extrinsic optical scattering loss in Photonic-crystal waveguides: role of fabrication disorder and photon group velocity}.
Phys. Rev. Lett. \textbf{94}, 033903 (2005)

\bibitem{Savona}
V. Savona.
\emph{Electromagnetic modes of a disordered photonic crystal}.
Phys. Rev. B \textbf{83}, 085301 (2011).

\bibitem{reviewMod}
P. Lodahl, S. Mahmoodian, and S. Stobbe.
\emph{Interfacing single photons and single quantum dots with photonic nanostructures}.
Rev. of Mod. Phys. \textbf{87}, 347 (2015).

\bibitem{Sheng}
P. Sheng.
\emph{Introduction to Wave Scattering, Localization, and Mesoscopic Phenomena}.
Academic Press, San Diego (1995).

\bibitem{Rossum}
M.~C.~W. van Rossum, and T.~M. Nieuwenhuizen.
\emph{Multiple scattering of classical waves: microscopy, mesoscopy, and diffusion}.
Rev. Mod. Phys. \textbf{71}, 313 (1999).

\bibitem{John}
S. John.
\emph{Strong localization of photons in certain disordered dielectric superlattices}.
Phys. Rev. Lett. \textbf{58}, 2486 (1987).

\bibitem{Gerace}
D. Gerace, and L. C. Andreani.
\emph{Effects of disorder on propagation losses and cavity q-factors in photonic crystal
slabs}.
Photonics Nanostruct. Fundam. Appl. \textbf{3}, 120 (2005).

\bibitem{C0}
P.~D. Garc\'{i}a, S. Stobbe, I.~S\"{o}llner, and P. Lodahl.
\emph{Nonuniversal intensity correlations in two-dimensional Anderson-localizing random medium}.
Phys. Rev. Lett. \textbf{109}, 253902 (2012).

\bibitem{Lifshitz}
I.~M. Lifshitz.
\emph{The energy spectrum of disordered systems}.
Adv. Phys. \textbf{13}, 483 (1964).

\bibitem{Huisman}
S.~R. Huisman, G. Ctistis, S. Stobbe, A.~P. Mosk, J.~L. Herek, A. Lagendijk, P. Lodahl, W.~L. Vos, and P.~W.~H. Pinkse.
\emph{Photonic-crystal waveguides with disorder: measurement of a band-edge tail in the density of states}.
Phys. Rev. B \textbf{86}, 155154 (2012).

\bibitem{TyrrestrupAPL}
H. Thyrrestrup, L. Sapienza, and P. Lodahl.
\emph{Extraction of the $\beta$-factor for single quantum dots coupled to a photonic-crystal waveguide}.
Appl. Phys. Lett. \textbf{96}, 231106 (2010).

\bibitem{mpb}
S.~G. Johnson and J.~D. Joannopoulos.
\emph{Block-iterative frequency-domain methods for Maxwell's equations in a planewave basis}.
Opt. Exp. \textbf{8}, 173 (2001).

\bibitem{Baba}
T. Baba.
\emph{Slow light in photonic crystals}.
Nat. Phot. \textbf{2}, 465 (2008).

\bibitem{Toke}
T. Lund-Hansen, S. Stobbe, B. Julsgaard, H. Thyrrestrup, T.~S\"{u}nner, M.~Kamp, A. Forchel, and P. Lodahl.
\emph{Experimental realization of highly efficient broadband coupling of single quantum dots to a photonic crystal waveguide}.
Phys. Rev. Lett. \textbf{101}, 113903 (2008).

\bibitem{Marta}
M. Arcari, I.~S\"{o}llner, A. Javadi, S.~L. Hansen, S. Mahmoodian, J. Liu, H. Thyrrestrup, E.~H. Lee, J.~D. Song, S. Stobbe, and P. Lodahl.
\emph{Near-unity coupling efficiency of a quantum emitter to a photonic crystal waveguide}.
Phys. Rev. Lett. \textbf{113}, 093603 (2014).

\bibitem{Smolka}
S. Smolka, H. Thyrrestrup, L. Sapienza, T.~B. Lehmann, K.~R. Rix, L.~S. Froufe-P\'{e}rez, P.~D. Garc\'{i}a, and P. Lodahl.
\emph{Probing statistical properties of Anderson localization with quantum emitters}.
New J. Phys. \textbf{13}, 063044 (2011).

\bibitem{quantifying}
P.~D. Garc\'{i}a, A. Javadi, H. Thyrrestrup, and P. Lodahl.
\emph{Quantifying the intrinsic amount of fabrication disorder in photonic-crystal waveguides from optical far-field intensity measurements}.
Appl. Phys. Lett. \textbf{102}, 031101 (2013).

\bibitem{red}
S.~E. Skipetrov, and J.~H. Page.
\emph{Red light for Anderson localization}.
New. J. Phys. \textbf{18}, 021001 (2016).

\bibitem{Chabanov}
A.~A. Chabanov, M. Stoytchev, and A.~Z. Genack.
\emph{Statistical signatures of photon localization}.
Nature \textbf{404}, 850 (2000).

\bibitem{Tigelen}
H. Hu, A. Strybulevych, J.~H. Page, S.~E. Skipetrov, and B.~A. van Tiggelen.
\emph{Localization of ultrasound in a three-dimensional elastic network}.
Nat. Phys. \textbf{4}, 945 (2008).

\bibitem{g}
E. Abrahams, P.~W. Anderson, D.~C. Licciardello, and T.~V. Ramakrishnam.
\emph{Scaling theory of localization: absence of quantum diffusion in two dimensions}.
Phys. Rev. Lett \textbf{42}, 673 (1979).

\bibitem{Noda}
S. Noda.
\emph{Seeking the ultimate nanolaser}.
Science \textbf{314}, 260 (2006).

\bibitem{Strong}
T. Yoshie, A. Scherer, J. Hendrickson, G. Khitrova, H.~M. Gibbs, G. Rupper, C. Ell, O.~B. Shchekin, and D.~G. Deppe.
\emph{Vacuum Rabi splitting with a single quantum dot in a photonic crystal nanocavity}.
Nature \textbf{434}, 200 (2004).

\bibitem{Qin1}
Q. Wang, S. Stobbe, and P. Lodahl.
\emph{Mapping the local density of optical states of a photonic crystal with single quantum dots}.
Phys. Rev. Lett. \textbf{107}, 167404 (2011).

\bibitem{kiraz}
A. Kiraz, P. Michler, C. Becher, B. Gayral, A. Imamo\~glu, L. Zhang, E.~Hu, W.~V. Schoenfeld, and P.~M. Petroff.
\emph{Cavity-quantum electrodynamics using a single InAs quantum dot in a microdisk structure}.
Appl. Phys. Lett. \textbf{78}, 3932 (2001).

\bibitem{Strong_Carminati}
A. Caz\'{e}, R. Pierrat, and R. Carminati.
\emph{Strong coupling to two-dimensional Anderson localized modes}.
Phys. Rev. Lett. \textbf{111}, 053901 (2013).

\bibitem{Strong_Tyrrestrup}
H. Thyrrestrup, S. Smolka, L. Sapienza, and P. Lodahl.
\emph{Statistical theory of a quantum emitter strongly coupled to Anderson-localized modes}.
Phys. Rev. Lett. \textbf{108}, 113901 (2012).

\bibitem{Wong}
J. Gao, S. Combrie, B. Liang, P. Schmitteckert, G. Lehoucq, S. Xavier, X. Xu, K. Busch, D.~L. Huffaker, A. De Rossi, and C.~W. Wong.
\emph{Strongly coupled slow-light polaritons in one-dimensional disordered localized states}.
Sci. Rep. \textbf{3}, 1994 (2013).

\bibitem{Javadi}
A. Javadi, S. Maibom, L. Sapienza, H. Thyrrestrup, P.~D. Garcia, and P. Lodahl.
\emph{Statistical measurements of quantum emitters coupled to Anderson-localized modes in disordered photonic-crystal waveguides}.
Opt. Exp. \textbf{22}, 30992 (2014)

\bibitem{Frank}
R. Frank, A. Lubatsch, and J. Kroha
\emph{Light transport and localization in diffusive random lasers}.
Journal of Optics A: Pure and Applied Optics \textbf{11}, 114012 (2009).

\bibitem{APL}
J.~K. Y.~H. Noh, M.~J. Rooks, G.~S. Solomon, F. Vollmer, and H. Cao.
\emph{Lasing in localized modes of a slow light photonic crystal waveguide}.
Appl. Phys. Lett. \textbf{98}, 241107 (2012).

\bibitem{Jin}
J. Liu, P.~D. Garc\'{i}a, S. Ek, N. Gregersen, T. Suhr, M. Schubert, J. M{\o}rk, S. Stobbe, and P. Lodahl.
\emph{Random nanolasing in the Anderson localized regime}.
Nat. Nanotech. \textbf{9}, 285 (2014).

\bibitem{Stano2012}
R. Stano, and P. Jacquod.
\emph{Suppression of interactions in multimode random lasers in the Anderson localized regime}.
Nat. Phot.. \textbf{7}, 66 (2013).

\bibitem{Cahotic_lasing1}
S. Mujumdar, V. T\"{u}rck, R. Torre, and D.~S. Wiersma.
\emph{Chaotic behavior of a random laser with static disorder}.
Phys. Rev. A \textbf{76}, 033807 (2007).

\bibitem{Cahotic_lasing2}
D.~S. Wiersma, S. Mujumdar, S. Cavalieri, S.~R. Torre, G.~L. Oppo, and S. Lepri.
\emph{Chaotic Behavor of Random Lasers.}
Chapter 10 in Tutorials in Complex Photonic Media. Eds., SPIE Press, Bellingham, WA, 277-299 (2009).

\bibitem{Riboli}
F. Riboli, N. Caselli, S. Vignolini, F. Intonti, K. Vynck, P. Barthelemy, A. Gerardino, L. Balet, L.~H. Li, A Fiore, M. Gurioli, and D.~S. Wiersma.
\emph{Engineering of light confinement in strongly scattering disordered media}.
Nat. Mater. \textbf{13}, 720 (2014).

\bibitem{Intonti}
V. Emiliani, F. Intonti, M. Cazayous, D.~S. Wiersma, M. Colocci, F. Aliev, and A. Lagendijk.
\textit{Near-field short range correlation in optical waves transmitted through random media}.
Phys. Rev. Lett. \textbf{90}, 250801 (2003).

\bibitem{Spasenovic}
M. Spasenovi\'{c}, D.~M. Beggs, P. Lalanne, T.~F. Krauss, and L. Kuipers.
\emph{Measuring the spatial extent of individual localized photonic states}.
Phys. Rev. B \textbf{86}, 155153 (2012).

\bibitem{Matteo}
G.~M. Conley, M. Burresi, F. Pratesi, K. Vynck, and D.~S. Wiersma.
\emph{Light Transport and Localization in Two-Dimensional Correlated Disorder}.
Phys. Rev. Lett. \textbf{112}, 143901 (2014).

\bibitem{correlations}
F. Riboli, F. Uccheddu, G. Monaco, N. Caselli, F. Intonti, M. Gurioli, and S.~E. Skipetrov.
\emph{Tailoring correlations of the local density of states in disordered photonic materials}.
Preprint can be checked online at https://arxiv.org/abs/1609.01975v1

\bibitem{Ricca}
R. Sapienza, P. Bondareff, R. Pierrat, B. Habert, R. Carminati, and N.~F. van Hulst.
\emph{Long-tail statistics of the Purcell factor in disordered media driven by near-field interactions}.
Phys. Rev. Lett. \textbf{106}, 163902 (2011).

\bibitem{Vos}
M.~D. Birowosuto, S.~E. Skipetrov, W.~L. Vos, and A.~P. Mosk,
\emph{Observation of spatial fluctuations of the local density of states in random photonic media}.
Phys. Rev. Lett. \textbf{105}, 013904 (2010)

\bibitem{Sandoghdar}
P.~V. Ruijgrok, R.W\"{u}est, A.~A. Rebane, A. Renn, and V. Sandoghdar
\emph{Spontaneous emission of a nanoscopic emitter in a strongly scattering disordered medium}.
Opt. Express \textbf{18}, 6360 (2010).

\bibitem{Lalanne2}
R. Faggiani, A. Baron, X. Zang, L. Lalouat, S.~A. Schulz, B. O'Regan, K. Vynck, B. Cluzel, F. de Fornel, T.~F. Krauss, and P. Lalanne.
\emph{Lower bound for the spatial extent of localized modes in photonic-crystal waveguides with small random imperfections}.
Scientific Reports \textbf{6}, 27037 (2016).

\bibitem{Lodahl2005}
P. Lodahl, A.~P. Mosk, and A. Lagendijk.
\emph{Spatial quantum correlations in multiple scattered light}.
Phys. Rev. Lett. \textbf{95}, 173901 (2005).

\bibitem{Saphiro}
B. Shapiro.
\emph{New type of intensity correlation in random media}.
Phys. Rev. Lett. \textbf{83}, 4733 (1999).

\bibitem{imaging}
B.~A. van Tiggelen, S.~E. Skipetrov.
\emph{Fluctuations of local density of states and C0 speckle correlations are equal}.
Phys. Rev. E \textbf{73}, 045601 (2006).

\bibitem{CO_Carminati}
A. Caz\'{e}, R. Pierrat, and R. Carminati.
\emph{Near-field interactions and nonuniversality in speckle patterns produced by a point source in a disordered medium}.
Phys. Rev. A \textbf{82}, 043823 (2010).

\bibitem{Vynk}
K. Vynck, M. Burresi, F. Riboli, and D.~S. Wiersma.
\emph{Photon management in two-dimensional disordered media}.
Nat. Mater. \textbf{11}, 1017 (2012).

\bibitem{Peretti}
R. Peretti, G. Gomard, Lo\"{i}c Lalouat, C. Seassal, E. Drouard.
\emph{Absorption control in pseudodisordered photonic-crystal thin films}.
Phys. Rev. A \textbf{88}, 053835 (2013).

\bibitem{Noda2}
A. Oskooi, P. A. Favuzzi, Y. Tanaka, H. Shigeta, Y. Kawakami, and S. Noda.
\emph{Partially disordered photonic-crystal thin films for enhanced and robust
photovoltaics}.
Appl. Phys. Lett. \textbf{100}, 181110 (2012).

\bibitem{Bertolotti}
J. Bertolotti, S. Gottardo, D.~S. Wiersma, M. Ghulinyan, and L. Pavesi
\emph{Optical necklace states in Anderson localized 1D systems}.
Phys. Rev. Lett. \textbf{94}, 113903 (2005).

\bibitem{Toninelli}
F. Sgrignuoli, G. Mazzamuto, N. Caselli, F. Intonti, F.~S. Cataliotti, M. Gurioli, and C. Toninelli
\emph{Necklace state hallmark in disordered 2D photonic systems}.
ACS Photonics, \textbf{2} (11), 1636 (2015).

\end{thebibliography}
\end{document}